\documentclass[floats,floatfix,showpacs,amssymb,prd,twocolumn,superscriptaddress,nofootinbib,nolongbibliography,reprint]{revtex4-2}

\usepackage{amssymb,amsmath,verbatim,mathtools,needspace,enumitem,etoolbox,graphicx,physics,microtype,afterpage,xspace,tabularx,lmodern,multirow,slashed}
\usepackage{gensymb}
\usepackage[normalem]{ulem}
\usepackage[dvipsnames, usenames]{xcolor}
\definecolor{linkcolor}{rgb}{0.0,0.3,0.5}
\usepackage[unicode, colorlinks=true, linkcolor=linkcolor, citecolor=linkcolor, filecolor=linkcolor, urlcolor=linkcolor, linktocpage, breaklinks]{hyperref}
\usepackage[all]{hypcap}
\usepackage[T1]{fontenc}
\usepackage[utf8]{inputenc}
\usepackage[usenames,dvipsnames]{xcolor}
\hypersetup{colorlinks=true,citecolor=romared,linkcolor=romared,urlcolor=romared}

\definecolor{romared}{RGB}{142,0,28}

\newcommand{\be}{\begin{equation}}
\newcommand{\ee}{\end{equation}}

\def\be{\begin{equation}}
\def\ee{\end{equation}}
\newcommand{\beq}{\begin{eqnarray}}
\newcommand{\eeq}{\end{eqnarray}}

\usepackage{makecell}
\usepackage{soul}

\newcolumntype{Y}{>{\centering\arraybackslash}X}

\begin{document}

\title{Curvature and dynamical spacetimes: can we peer into the quantum regime?}

\author{Vitor Cardoso} 
\affiliation{Niels Bohr International Academy, Niels Bohr Institute, Blegdamsvej 17, 2100 Copenhagen, Denmark}
\affiliation{CENTRA, Departamento de F\'{\i}sica, Instituto Superior T\'ecnico -- IST, Universidade de Lisboa -- UL, Avenida Rovisco Pais 1, 1049-001 Lisboa, Portugal}

\author{David Hilditch} 
\affiliation{CENTRA, Departamento de F\'{\i}sica, Instituto Superior T\'ecnico -- IST, Universidade de Lisboa -- UL, Avenida Rovisco Pais 1, 1049-001 Lisboa, Portugal}
\author{Krinio Marouda} 
\affiliation{CENTRA, Departamento de F\'{\i}sica, Instituto Superior T\'ecnico -- IST, Universidade de Lisboa -- UL, Avenida Rovisco Pais 1, 1049-001 Lisboa, Portugal}

\author{Jos\'e Nat\'ario}
\affiliation{CAMGSD, Departamento de Matem\'{a}tica, Instituto Superior T\'ecnico -- IST, Universidade de Lisboa -- UL, Avenida Rovisco Pais 1, 1049-001, Lisboa, Portugal}

\author{Ulrich Sperhake}
\affiliation{Department of Applied Mathematics and Theoretical Physics, Centre for
Mathematical Sciences, University of Cambridge, Wilberforce Road, Cambridge
CB3 0WA, United Kingdom}

\date{\today}

\begin{abstract}
Stationary compact astrophysical objects such as black holes and neutron stars behave as classical systems from the gravitational point of view. Their (observable) curvature is everywhere ``small''. Here we investigate whether mergers of such objects, or other strongly dynamical spacetimes such as collapsing configurations, may probe the strong-curvature regime of General Relativity. Our results indicate that dynamical black hole spacetimes always result in a modest increase $\sim 3$ in the Kretschmann scalar, relative to the stationary state. Our results show
that the Kretschmann scalar can dynamically increase by orders of magnitude, during the gravitational collapse of scalar fields, and that the (normalized) peak curvature does {\it not} correspond to that of the critical solution. Nevertheless, without fine tuning of initial data, this increase lies far below that needed to render quantum-gravity corrections important.
\end{abstract}

\maketitle

\noindent {\bf \em Introduction.} The advent of gravitational-wave (GW) astronomy~\cite{LIGOScientific:2016aoc,Abbott:2020niy} and of very long baseline interferometry~\cite{EventHorizonTelescope:2019dse,GRAVITY:2020gka} opened
exciting new windows to the invisible Universe.
Compact objects, in particular neutron stars and black holes (BHs), play a unique role in the endeavor to test our understanding of general relativity (GR) and in the search for new physics~\cite{Berti:2015itd,Barack:2018yly,Berti:2018cxi,Berti:2018vdi,Cardoso:2019rvt,Bertone:2018krk,Brito:2015oca}. 

According to the singularity theorems~\cite{Penrose:1964wq,Penrose:1969pc}, classical GR must fail in BH interiors. Quantum mechanics in BH spacetimes also leads to puzzling consequences, such as the information paradox~\cite{Unruh:2017uaw,Mathur:2005zp,Giddings:2017mym}.
It is tempting to conjecture that a theory of quantum gravity will resolve these issues, but the scale and nature of quantum gravity corrections to BH spacetimes is unknown.

In this paper we ask a simple but outlandish question: can dynamical, astrophysical processes probe the quantum gravity regime? 

The inclusion of backreaction of quantum fluctuations is a delicate problem in General Relativity~\cite{Birrell:1982ix}. Working at semi-classical level, the renormalized 1-loop effective action admits a low-curvature expansion of the form~\cite{Birrell:1982ix,Parker:2008,Burgess:2003jk}
\beq
S_{\rm eff}&=&\int d^4x \sqrt{-g}\left(-\Lambda_{\rm eff}+\frac{R}{16\pi G_{\rm eff}}+{\cal L}_{\rm eff}^{(1)}+\ldots\right),\\
{\cal L}_{\rm eff}^{(1)}&=&{M_{\rm P}}^2\left(a_1 R^2+a_2R_{\alpha\beta}R^{\alpha\beta}+a_3R_{\alpha\beta\gamma\delta}R^{\alpha\beta\gamma\delta}\right),
\eeq
where $\Lambda_{\rm eff}\,,G_{\rm eff}$ are the effective cosmological and Newton's constant, respectively, and $M_{\rm P}$ is the Planck mass (we work with geometric units  $c=1=G_{\rm eff}$ throughout).
The precise values of the constants $a_i$ are not necessary for us here, but they can be calculated exactly in a semi-classical framework. The important aspect is that higher-curvature terms are expected to be present generically; variants of this argument come from effective field theory approaches. The most general theory without new degrees of freedom, compatible with observations, and obeying some basic principles, can be shown to be Einstein's theory corrected by higher-order derivative terms~\cite{Birrell:1982ix,Parker:2008,Burgess:2003jk,Endlich:2017tqa}.

The above provides a general framework to look for quantum gravitational effects.
Setups for which the Kretschmann or other higher-derivative invariants are too large are not described by small corrections to the classical theory: one is then in the quantum regime. A similar motivation is behind recent studies of quantum gravitational anomalies and their impact on photons from coalescences of compact objects~\cite{Agullo:2016lkj,delRio:2020cmv}, or studies of gravity with higher-order corrections~\cite{Cardoso:2018ptl}.

Unfortunately, the (lowest-order) corrections relative to the classical equations of motion are expected to scale as $\left(M_{\rm P}/M\right)^2 \sim 10^{-76}$ or smaller, with $M$ the mass of the macroscopic object under study (say, a neutron star or stellar mass BH). For {\it equilibrium} compact configurations, these quantum corrections are indeed small: consider a neutron star of mass $M$ and radius $r\sim 6M$; the Kretschmann scalar -- Eq.~\eqref{eq:K1} -- is $K_1\lesssim 0.1 M^{-4}$, whereas the Ricci scalar at the surface is $R\sim M/r^3=0.04M^{-2}$. Thus, in such an equilibrium scenario, higher order curvature corrections are expected to be around 76 orders of magnitude smaller than the classical terms.

In the absence of the ultimate theory of quantum gravity, we ask the following question: can higher curvature terms {\it ever} become important during a {\it dynamical} evolution? Take a neutron star collapsing to a non-spinning BH: we know that the BH interior harbours strong-curvature regions; is it possible that outward propagating photons or gravitons had access to ``unnaturally large'' curvature regions? This question is related, but not identical, to claims on cosmic censorship violations~\cite{Emparan:2020vyf}. In other words, we are not so much concerned with curvatures reaching arbitrarily large values but rather with how much larger than their stationary values they can become during a generic dynamical evolution (and in particular whether they can become Planckian).

\noindent
{\bf \em Quantities of interest.}
There are five quadratic invariants constructed from the Riemann and the Weyl tensors~\cite{Cherubini:2002gen} (there are also cubic invariants, but we will not consider them here): the Kretschmann invariant, the Chern-Pontryagin invariant, the Euler invariant, and the first and second Weyl invariants. Three of these can be computed from the others once the energy momentum tensor is known. Therefore, we focus solely on two invariants, the 
Kretschmann scalar
\begin{equation}
K_1 \equiv R_{\alpha\beta\gamma\delta} R^{\alpha\beta\gamma\delta}
\,,
\label{eq:K1}
\end{equation}
and the Chern-Pontryagin invariant
\begin{equation}
K_2 \equiv ^{\,\,\star\!\!}RR=\frac{1}{2}R_{\alpha\beta\gamma\delta} \epsilon^{\alpha\beta\mu\nu}R^{\gamma\delta}_{\quad\mu\nu} \,.
\label{eq:K2}
\end{equation}
Note that $K_2$ is necessarily zero for spherically symmetric spacetimes, as the antipodal map is an orientation-reversing isometry.

\begin{figure}[!t]
    \centering
    \includegraphics[width=9cm]{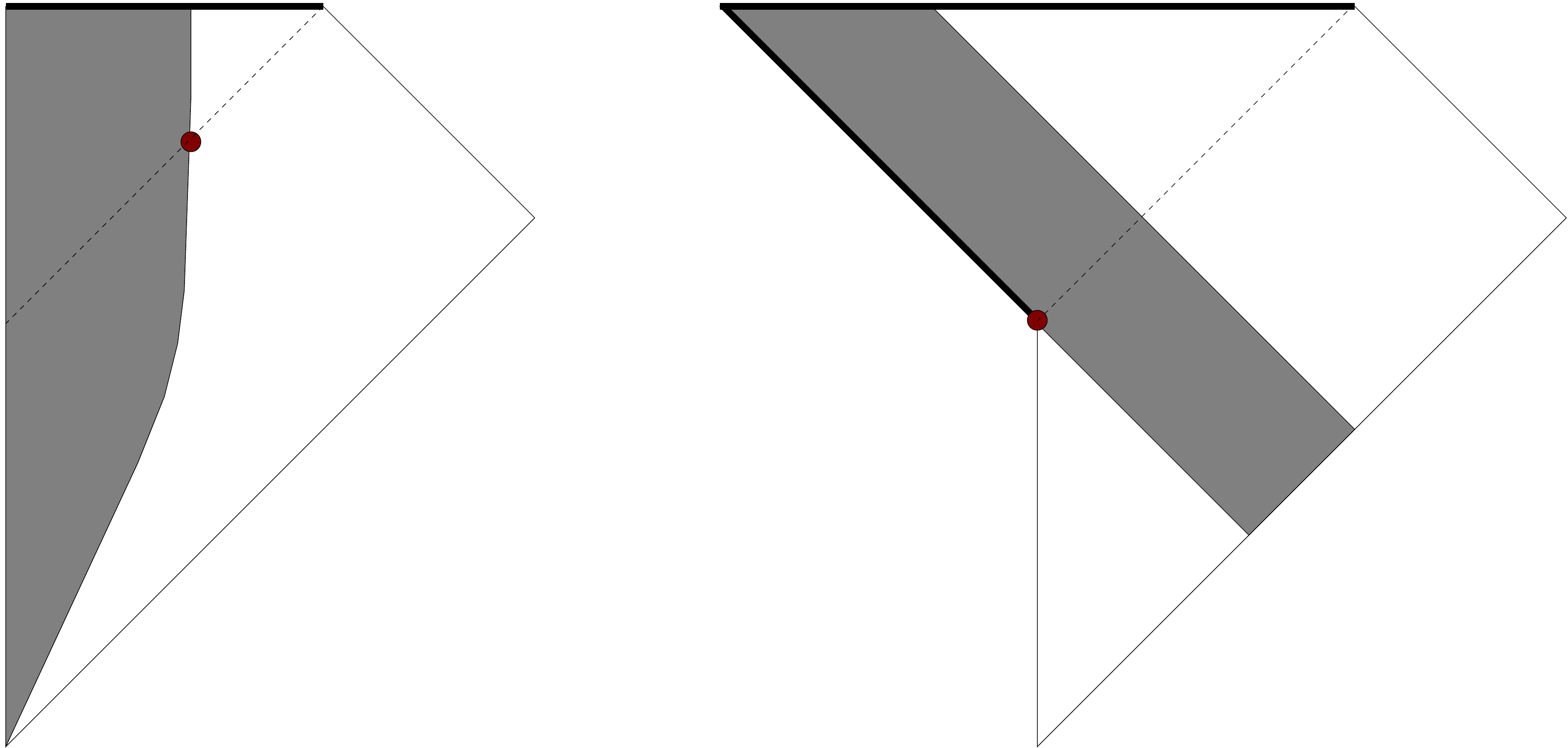}
    \caption{Penrose diagrams for the Oppenheimer-Snyder and the critical null dust collapses. The shaded regions correspond to the dust, the dashed lines to the event horizons, and the thick lines to the singularities. The red dots mark the points where the Kretschmann invariant $K_1$ is maximal along the horizon (infinite in the case of the critical Vaidya solution).}
    \label{fig:Penrose}
\end{figure}

\noindent
{\bf \em Oppenheimer-Snyder collapse.}
Arguably, the simplest model of a gravitational collapse is that of Oppenheimer-Snyder, where a ball of dust with uniform density collapses to a Schwarzschild BH. The metric inside the dust ball can therefore be written as
\begin{equation}
ds^2 = - d\tau^2 + a^2(\tau) (d\rho^2 + \rho^2 d \Omega^2) \, ,
\end{equation}
where $d \Omega^2$ is the line element of the unit $2$-sphere and we assumed, for simplicity, the metric to be spatially flat. The metric outside the dust ball is, of course, the Schwarzschild metric
\begin{equation}
ds^2 = -\left(1 - \frac{2M}{r}\right) dt^2 + \left(1 - \frac{2M}{r}\right)^{-1} dr^2 + r^2 d \Omega^2 \, ,
\end{equation}
where $M$ is the mass of the final BH. At the boundary of the dust ball we have \cite{Natario21}
\begin{equation}
\mu = \frac{3M}{4\pi r^3} \, ,
\end{equation}
where $\mu$ is the ball's density. As the Weyl tensor vanishes inside the dust ball, $K_1= \frac53 (8 \pi \mu)^2$, corresponding to
\begin{equation}
K_1 = \frac{60M^2}{r^6}
\end{equation}
at the ball's boundary.
Since the Kretschmann invariant $K_1$ is constant along constant $\tau$ hypersurfaces, and increasing with $\tau$, the events in the dust ball with largest $K_1$ visible from the exterior correspond to the intersection of the dust ball's boundary with the event horizon (see Fig.~\ref{fig:Penrose}), where 
\begin{equation}
K_1 = \frac{15}{16M^4} \, .
\end{equation}
This is slightly larger (by a factor of $5/4$) than the maximum value observable in the Schwarzschild exterior, where
\begin{equation}
K_1 = \frac{48M^2}{r^6} < \frac{3}{4M^4} \, .\label{K1_Sch}
\end{equation}
It is interesting to note that the nature of the curvature inside and outside the dust ball is completely different: whereas inside the Weyl tensor vanishes and there is only Ricci curvature, outside it is the Ricci tensor that vanishes, leaving only the Weyl curvature.

It is possible to obtain much larger values of the Kretschmann scalar by considering inhomogeneous dust collapses. In fact, it is well known that such collapses can even produce naked singularities, corresponding to shell crossings or shell focusing \cite{EardleySmarr:1979, Christodoulou:1984}. If one is careful to cover these singularities with the event horizon \cite{JoshiMalafarina:2015}, arbitrarily large (but finite) values of $K_1$ can be observed from the exterior; however, this requires fine-tuning, and such collapses are quite dissimilar to realistic astrophysical collapses. A straightforward example of this can be seen in the limiting case of null dust collapses, which we now briefly discuss.

\noindent {\bf \em Null radiation collapse.}
Another simple model of collapse can be obtained form the Vaidya metric \cite{Vaidya:1951}, given by
\begin{equation}
ds^2 = -\left(1 - \frac{2M(v)}{r}\right) dv^2 + 2 dv dr + r^2 d \Omega^2,
\end{equation}
where $M(v)$ is an arbitrary function of the advanced time $v$. This metric corresponds to the energy-momentum tensor
\begin{equation}
T = \frac{1}{4\pi r^2} \frac{dM}{dv} dv\otimes dv \, ,
\end{equation}
describing a null dust propagating along the incoming radial null geodesics. Choosing $M(v)$ to vary from $0$ to some final value $M_1$ on some interval $v \in [0,v_1]$, and constant outside this interval, provides a simple model of BH formation, as the metric coincides with the Minkowski metric for $v<0$ and with the Schwarzschild metric of mass $M_1$ for $v>v_1$. The Kretschmann scalar is still given by equation \eqref{K1_Sch}, and so we see that we can obtain arbitrarily large curvatures outside the event horizon if $M(v)>0$ near $r=0$. It turns out that by fine-tuning the free function $M(v)$ it is possible to arrange for the Penrose diagram of the Vaydia solution to be as depicted in Fig.~\ref{fig:Penrose} \cite{HiscockWilliamsEardley:1982, Kuroda:1984, Joshi:1997, FayosTorres:2010} (as an example, if $M(v)$ is taken to be of the form $M(v)=\mu v$ for $v\in[0,v_1]$ then the fine-tuning amounts to choosing $8M_1/v_1=1+\sqrt{1-\mu^2}$). This diagram represents a critical solution, separating naked singularities from BHs; for these solutions, the event horizon emanates from the first singularity, and so $K_1$ attains arbitrarily large values on the event horizon (in the example above, $K_1 = C/v^4$ along the event horizon, where $C=C(\mu)$ is a positive constant). Small perturbations of these solutions lead to BHs whose event horizons no longer emanate from the first singularity, and so we can have arbitrarily large (but finite) values of $K_1$ visible from the exterior. However, these perturbations are very particular solutions, quite distant from any conceivable astrophysical scenario. They would correspond to initial data either already displaying large curvatures, or carefully fine-tuned to produce them via focusing.

\noindent {\bf \em Massless minimally coupled scalar field collapse.} 
Another classical setup of collapse in spherical symmetry is the numerical study of BH formation due to a self-gravitating scalar field minimally coupled to gravity. We will use this as a proxy for gravitational collapse of stars to BHs in astrophysical setups. We use horizon-penetrating coordinates, namely Kerr-Schild coordinates, which allow probing the formation of the apparent horizon and also to study the behaviour of the curvature invariants throughout the whole evolution, before and after the horizon forms. The metric is built using ingoing and outgoing null vectors and their associated covectors:
\begin{align*}
  \xi= \partial_t + C_+\partial_r\,, \quad
  \text{\underbar{$\xi$}}=\partial_t + C_-\partial_r \, ,
\end{align*}
\begin{align*}
  \eta=-C_+ dt +dr\,, \quad
     \text{\underbar{$\eta$}}=C_-dt-dr \, .
\end{align*}
Here, $C_+$ and $C_-$ control the local behaviour of the lightcones in these coordinates. Due to spherical symmetry, all the evolved variables in this section are functions of $(t,r)$ only. The spherically symmetric ansatz for the metric using these null covectors is the following:
  \begin{align}
  ds^2=-\frac{e^{\delta}}{C_+-C_-}\left(\eta \otimes\text{\underbar{$\eta$}} +\text{\underbar{$\eta$}}\otimes \eta\right) +r^2 d\Omega^2 \,,
  \end{align}
  where $\delta$ is associated to the determinant of the metric in the $(t,r)$ plane. We will be using the Kerr-Schild gauge~\cite{Bhattacharyya_2021} with $r$ areal radius and $C_-=-1$, and work with the renormalized variable 
\begin{align}
    \tilde{C}_+=\frac{2 C_+}{C_++1} 
\end{align}
 In other words, the ingoing characteristic speed has been set to $C_-=-1$ everywhere in spacetime, while the radial coordinate has been taken to be the areal radius.
 \begin{figure}[!t]
    \centering
    \includegraphics[width=9cm]{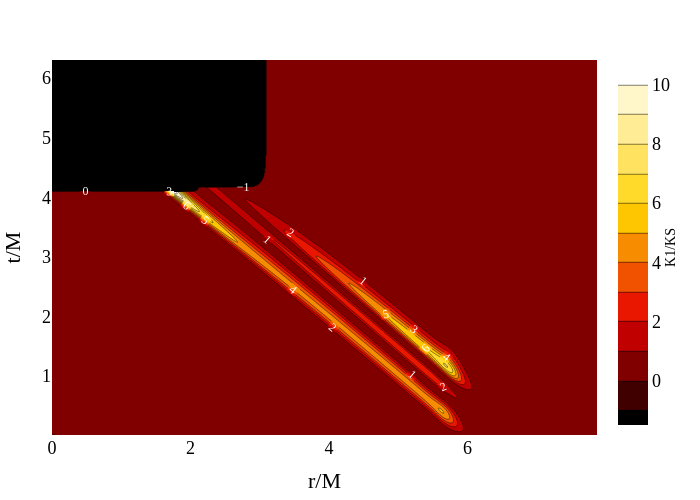}
    \caption{Contour of the Kretschmann invariant $K_1$ in the $(t,r)$ plane in units of the relevant invariant $K_S$ in Eq.~\eqref{eq:KS}, of a static Schwarzschild with mass equal to the mass of the apparent horizon of a newly formed BH, $M_{\text{AH}}=3.17$. The invariant $K_1$ grows during the evolution, with the maximum happening close to the initial location of the apparent horizon.}
    \label{fig:figure_1}
\end{figure}
 \begin{figure*}[!t]
    \centering
    \includegraphics[width=0.45\textwidth]{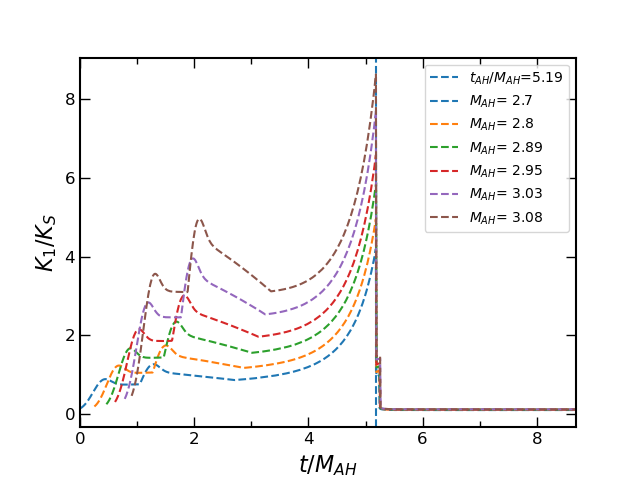}
    \includegraphics[width=0.45\textwidth]{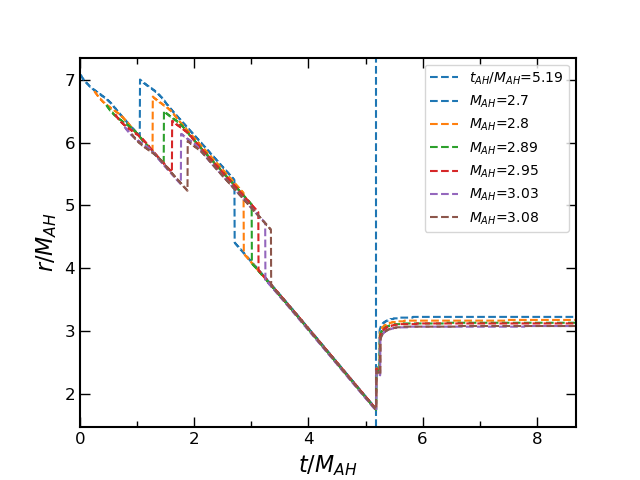}
    \caption{{\bf Left Panel:} evolution of the maximum of the Kretschmann invariant $K_1$ normalized by $K_S$ (see Eq.~\eqref{eq:KS}). All the plots have been shifted in time so that the apparent horizon formation occurs at $t_{\rm AH}=5.2M_{\rm AH}$. The initial pulse is centered at $R_0=20$, with $\sigma=2$. {\bf Right Panel:} location of the maximum of $K_1$ during the same evolution. Note that the location of the maximum jumps in the beginning because some other local maximum takes over as the global maximum. However, close to the time of the apparent horizon formation, the global maximum is at some radius smaller than the radius of the apparent horizon that forms later.}
    \label{fig:figure_2}
\end{figure*}
 
 The field equations with a stress energy tensor of a massless minimally coupled scalar field $\psi$ in the matter sector reduce to two evolution equations
\begin{align}
    \nabla_{\underline{\xi}}\delta&= -4 \pi  r T_{\underline{\xi}\underline{\xi}} \, , \\
    \nabla_{\underline{\xi}}\left(re^{-\delta}\tilde{C}_+\right)&=4 \pi  r^2e^{-\delta} T_{\text{\underbar{$\xi$}\underbar{$\xi$}}}-1 \, ,
\end{align}
and one constraint equation
\begin{align}
\label{Eq:constraint}
   \frac{\partial_r \left(re^{-\delta}\tilde{C}_+\right)-1}{\tilde{C}_+-2}=2 \pi r^2e^{-\delta} \left(\tilde{C}_+ T_{\underline{\xi}\underline{\xi}}-(\tilde{C}_+-2) T_{\xi \xi
   }\right) \, .
\end{align}
We solve these field equations for the variables~$\tilde{C}_+,\delta$ together with the wave equation for the scalar field, $\Box_g\psi=0$, written as a first order system on the variables~$\Pi=\partial_t\psi$ and~$\Phi=\partial_r\psi$. We use a Runge-Kutta with method of lines, second order finite differencing, and Kreiss-Oliger dissipation.

The ADM mass in these coordinates is computed as
\begin{equation}
M_{\textrm{ADM}}=\underset{R\to\infty}{\text{lim}}\left[r\left(1-\frac{e^{-\frac{\delta}{2}}}{\sqrt{2-\tilde{C}_+}}\right)\right] ,
\end{equation}
and is independent of~$T$ throughout the evolution, since the spacelike slices capture both the strong field region and the scalar radiation. The horizon mass is given by 
\begin{equation}
M_{\textrm{AH}}=\sqrt{\frac{A_{\rm AH}}{16\pi}} \, ,
\end{equation}
where $A_{\rm AH}$ is the area of the apparent horizon. Throughout we will discuss results obtained via apparent horizon properties. We have calculated also the event horizon location dynamically in a subset of cases, and find that our conclusions are not affected qualitatively.

We choose the scalar field in the initial~$(t=0)$ slice to be a localised Gaussian of amplitude~$P$ and standard deviation~$\sigma$ centered at~$r=r_0$, 
\begin{align}
    \psi(0,r)=\frac{P}{\sqrt{2\pi \sigma}} \left(e^{-\left(\frac{r-r_0}{\sigma}\right)^2} +e^{-\left(\frac{r+r_0}{\sigma}\right)^2} \right),
\end{align}
while all the other free functions are set to zero, and solve for $\tilde{C}_+(0,r)$ using the constraint equation~\eqref{Eq:constraint}. There is no BH in our initial slice because using the above initial profiles for the grid functions and assuming regularity at the origin, $\tilde{C}_+(0,0)=1$, the equation admits a solution of the form
\begin{align*}
    \tilde{C}_+(0,r)=\frac{1}{r} e^{-4\pi \int_0^r \tilde{r} \Phi(0,\tilde{r})^2 d\tilde{r}}\left[1+e^{4\pi\int_0^\zeta \tilde{r} \Phi(0,\tilde{r})^2 d\tilde{r}}\right]>0 \, .
\end{align*} 
where $\zeta$ is an integration constant.
 For sufficiently large initial amplitude, however, an apparent horizon forms during the evolution, which in these coordinates simply corresponds to the condition~$\tilde{C}_+ = 0$. In fact, the expansion of the congruence of radial outgoing null geodesics is computed as 
\begin{equation}
    \Theta=\frac{2C_+}{r}\,,
\end{equation} 
in these coordinates. After an apparent horizon forms, we perform BH excision a few time steps later, and keep on evolving the domain of outer communications.

To understand how the dynamics drives the curvature invariants, we normalize our results by the Kretschmann scalar of the newly formed BH, i.e., expression \eqref{K1_Sch} evaluated at the apparent horizon,
\be
K_{\text{S}}=\frac{3}{4M_{\rm AH}^4}\,.\label{eq:KS}
\ee
In this way, we avoid getting large curvature invariants as an artifact of ``small'' initial conditions that led to a small BH: we are interested in astrophysical setups, and therefore would like to learn if large gradients are possible even when the final object is a stellar-mass BH.

\begin{figure}[!h]
    \centering
    \includegraphics[width=0.5\textwidth]{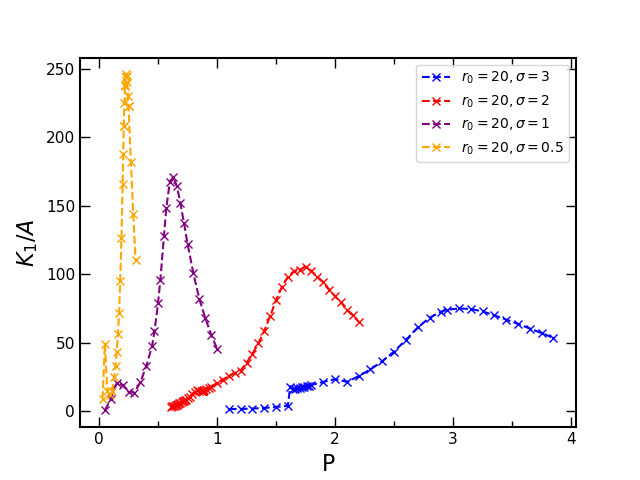}
    \caption{Global maximum of $K_1$ for the scalar field collapse, in units of $A=\max{\left(\left. K_{\text{max}} \right|_{t=0}, K_S\right)}$, shown for four series of evolutions of pulses centered at $r_0=20$, with width $\sigma=0.5,\,1,\,2,\,3$, respectively, and varying P. The discontinuity observed for $\sigma=3$ is due to the outgoing scalar pulse being trapped in the photon sphere of the BH that forms in early stages of the evolution, and contributes to the global maximum. The global peaks of these series of evolutions follow an empirical law calculated as
    $\max{K_1}=\alpha \left( \frac{r_0}{\sigma} \right)^{\gamma}$, for small $\sigma$, where $\alpha=33\pm 3$ and $\gamma=0.6\pm 0.1$.
    }
    \label{fig:figure_3}
\end{figure}

Our results are summarized in Figs.~\ref{fig:figure_1}--\ref{fig:figure_3}.
After the apparent horizon forms, the peak of $K_1$ occurs close to the AH/EH for ``small'' amplitude of initial data, while for ``stronger'' data the maximum can also be located away from the horizon, since there is an outgoing pulse of scalar radiation.

For the range of initial data that we have studied here, we observe that~$K_{\text{max}}/K_{\text{S}}$ is always~$\lesssim10^3$ (see Fig.~\ref{fig:figure_2}). This is a much greater increase than in the analytic example of homogeneous dust collapse discussed above.
We find that for some configurations the maximum curvature $K_1$ may arise from early time evolution of data which itself has a large curvature. Although we suspect that this problem can be circumvented by prescribing initial data ``further in the past,'' we decided to be conservative and normalize our results to $K_1/A$, with $A=\max{\left(\left. K_{\text{max}} \right|_{t=0}, K_S\right)}$. The results are shown in Fig.~\ref{fig:figure_3}.

For a fixed radius $r_0$ and width $\sigma$, we find that the curvature $K_1$ peaks at a finite value of the amplitude $P$, as depicted in Fig.~\ref{fig:figure_3}. Our results are well described by
\be
\frac{\max{K_1}}{A}=\alpha \left( \frac{r_0}{\sigma}\right)^{\gamma}\,,\label{max_K1_collapse}
\ee
for small $\sigma$, where $\alpha=33\pm 3$ and $\gamma=0.6\pm 0.1$.

For a very crude estimate of what this means, arrange all neutrons in a neutron star in a shell of thickness $\sigma$ close to the star radius, so as to maximize \eqref{max_K1_collapse}. Now let the configuration collapse, and find
\be
\frac{\max{K_1}}{A}\lesssim 10^{13}\left(\frac{r_0}{10^4\,{\rm m}}\right)^{0.6} \left( \frac{10^{-15}\,{\rm m}}{\sigma}\right)^{0.6}\,.
\ee
Even in such a highly idealized (and impossible, since the neutron star material could not possibly be arranged in such a shell) configuration, one is far below the 76 orders of magnitude necessary to reach Planckian curvatures.

In those cases where we ran our event horizon finder, we found that it forms slightly earlier than the apparent horizon, and that it covers a portion of approximately $50 \% $ of the observed Kretschmann peaks in Fig.~\ref{fig:figure_2}.

\noindent {\bf \em BH binary merger.}
As an independent scenario, we have also investigated the
time evolution of the curvature scalars in the inspiral
and merger of a BH binary. We expect the
curvature dynamics to be most pronounced around merger and
therefore select for our study the relatively short
(about 2 orbits) non-spinning, equal-mass binary
labeled R1 in Table I of Ref.~\cite{Baker:2006yw}:
each BH has a bare mass $m=0.483$ and they start at
$x=\pm 3.257$ with tangential momentum $P_y=\pm 0.133$
resulting in a total BH mass of $M=1.01$ (all in code units).

We simulate this binary with the {\sc Lean} code
\cite{Sperhake:2006cy} which is based on the
{\sc Cactus} computational toolkit \cite{Goodale2002}
and employs {\sc Carpet} \cite{Schnetter:2003rb} for mesh refinement.
The Einstein equations are evolved using the
Baumgarte-Shapiro-Shibata-Nakamura-Oohara-Kojima
(BSSNOK) formalism
\cite{Nakamura:1987zz,Shibata:1995we,Baumgarte:1998te}
with the moving-puncture gauge
\cite{Campanelli:2005dd,Baker:2005vv}
and apparent horizons are computed with \textsc{AHFinderDirect}
\cite{Thornburg:1995cp,Thornburg:2003sf}.

In vacuum, the Riemann and Weyl tensor are identical
and we compute the curvature scalars $K_1$ and $K_2$
from the electric and magnetic parts of the Weyl tensor
\begin{equation}
  E_{\alpha\beta} = C_{\alpha \mu \beta\nu} n^{\mu}n^{\nu}\,,
  ~~~~~
  B_{\alpha\beta} = \frac{1}{2}
  \epsilon_{\alpha\mu}{}^{\rho \sigma}
  C_{\rho\sigma\beta\nu}n^{\mu}n^{\nu}\,,\end{equation}
where $n^{\mu}$ denotes the timelike unit normal field;
cf.~Refs.~\cite{Friedrich:1996hq,Sperhake:2006cy} for
more details. In vacuum, Eqs.~(\ref{eq:K1}) and (\ref{eq:K2})
expressed in terms of the electric and magnetic parts
become
\begin{eqnarray}
  K_1 &=& 8(E^{mn}E_{mn}-B^{mn}B_{mn})\,,
  \label{eq:K1EB}
  \\[10pt]
  K_2 &=& 16 E^{mn}B_{mn}\,,
\end{eqnarray}
where we have switched from (Greek) spacetime indices
to (Latin) spatial indices since the electric and magnetic
parts of the Weyl tensor are by construction purely spatial
tensors, $E_{\mu\nu}n^{\nu} = 0 = B_{\mu\nu}n^{\nu}$.

Here, we are interested in the maximal values of the
curvature scalars that are realized {\it outside} the
BH horizons. This exclusion of the horizon's interior
encounters three practical difficulties: (i) Failure
to find an apparent horizon does not necessarily imply
absence of a horizon. (ii) Points outside the
apparent horizon may be inside the event horizon. (iii)
For non-spherical horizons, it is technically challenging
to determine if a given grid point is inside the horizon.

The first difficulty is mitigated by the high reliability
of {\sc AHFinderDirect}; in every simulation, the AH finder fails to determine a horizon at exactly one time
step around merger; we ignore this time step in our analysis.
The second difficulty can only be overcome by computing
event horizons which, however, is a highly complicated
task (see e.g.~\cite{Diener:2003jc}) and which we leave for future studies. Finally, we address
the third challenge by evaluating two estimates
of the maximum curvature, one by excising from
the calculation a sphere with the {\it maximal}
horizon radius and a second by excising instead
a sphere with the {\it minimal} horizon radius.
Both these radii are readily provided by
{\sc AHFinderDirect} together with the
centroid of the horizon. The former gives us
a conservative lower estimate for the maximal
curvature (since we may have discarded legitimate points outside the exact horizon)
while the latter gives us a strict upper limit
(every point ignored in this calculation is
definitely inside the apparent horizon and, thus, also
inside the event horizon). In the following,
we refer to these two methods as the $r_{\rm min}$
and $r_{\rm max}$ methods.

\begin{figure}
\includegraphics[width=0.5\textwidth]{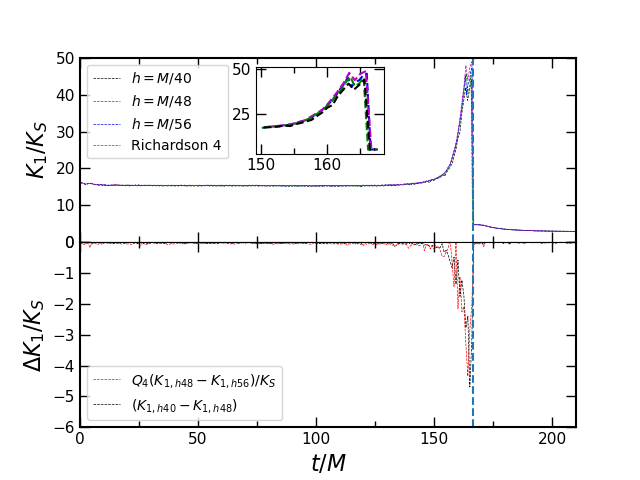}
\caption{
The curvature scalar $K_1$, maximized over space,
is plotted as a function of time for a $\sim 2$ orbit inspiral and merger of two non-spinning, equal-mass BHs.
The upper panel shows the resulting $\max(K_1)$ 
for the three grid resolutions
together with a fourth-order Richardson extrapolation.
The bottom panel shows the differences
between low and medium, as well as between medium
and high resolution. The latter is scaled
with a factor $Q_4=2.333$ expected for
fourth-order convergence. The vertical dotted line
around $t/M\approx 166.5$ marks the first occurence of a
common apparent horizon. All curvature estimates are
normalized to the Schwarzschild value on the horizon, Eq.~\eqref{eq:KS}.
}
\label{fig:k1_conv}
\end{figure}

We calibrate the numerical accuracy of our
calculations with a convergence analysis obtained
from three simulations using a grid setup
\begin{equation}
  \left\{
  (208,128,72,24,12,6)\times
  (1.5,0.75),~h
  \right\}\,.
  \nonumber
\end{equation}
That is, we have two inner refinement levels, each consisting
of two boxes of ``radius'' 0.75 and 1.5 centered around
either BH, and 6 outer levels of ``radius'' 6, 12, 24, 72,
128 and 208 (all in units of total BH mass $M$)
centered on the origin. The grid spacing is
$h$ on the innermost level and increases by a factor 2
on each level further out. In Fig.~\ref{fig:k1_conv},
we plot the resulting maximal curvature obtained
for the $r_{\rm max}$ method in units of the
Kretschmann scalar $K_S$ on the horizon of
a Schwarzschild BH of mass $M$, Eq\eqref{eq:KS}. In order to reduce
high-frequency noise in the convergence analysis,
we compare in the lower panel of the figure the differences
between our finite-resolution results using a 10 point
running average of the function $\max (K_1)$. This
high-frequency noise arises from the lego sphere nature of
the region we discard from the evaluation of $K_1$;
as the BHs move across the domain, ``optimal''
grid points can cross the horizon, resulting in
a sudden drop or jump in $\max (K_1)$. The averaging
procedure does not significantly affect the
resulting convergence estimate, but greatly enhances
the readability of the figure; note that only differences
in $\max(K_1)$ have been averaged in Fig.~\ref{fig:k1_conv}, but not $\max(K_1)$ itself. 

The bottom panel
demonstrates convergence close to fourth order which
we employ in the Richardson extrapolation displayed
in the upper panel. Based on this extrapolation, we
obtain a discretization error of $4\,\%$ around
merger and below $1\,\%$ throughout inspiral and ringdown.
\begin{figure}
\includegraphics[width=0.5\textwidth]{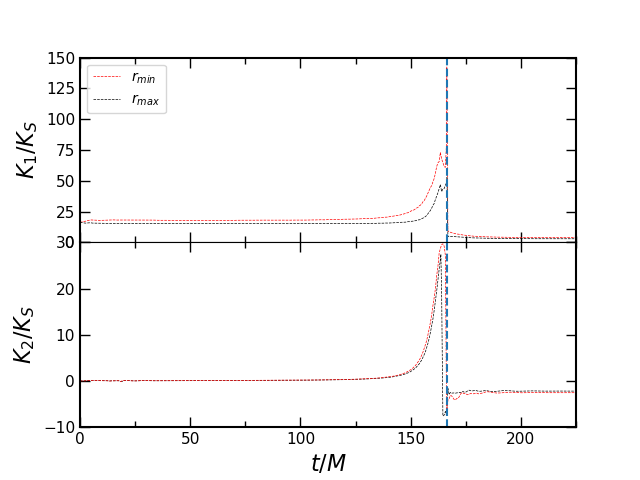}
\caption{
The curvature scalars $K_1$ and $K_2$, maximized in modulus over
space, are plotted as functions of time for a $\sim 2$ orbit inspiral
and merger of two non-spinning, equal-mass BHs. In each panel,
the black curve
shows the result obtained for the $r_{\rm max}$
method and the red curve shows the results for the
$r_{\rm min}$ method. The vertical dotted line at
$t/M \approx 166.5$ shows the time of first identification
of a common apparent horizon.
}
\label{fig:max_k1k2}
\end{figure}

The results for $K_1$ and $K_2$ extracted from our
BH simulations are displayed in Fig.~\ref{fig:max_k1k2}. For each scalar, we plot in this
figure the maximum value obtained outside our $r_{\rm max}$
(black curves) and $r_{\rm min}$ (red curves) approximation of the
apparent horizon. During the inspiral phase up to about
$t=150\,M$, we see that $K_1$ remains close to the value
$K_1=16~K_{\rm S}$ expected on the horizon of a Schwarzschild
BH of mass $M/2$ while $K_2$ vanishes as
expected for a non-spinning BH. Around merger,
marked by the vertical dotted line in the figure, both
curvature scalars rapidly increase to $K_1\approx 75~K_{\rm S}$
and $K_2\approx 30~K_{\rm S}$, respectively before dropping
to the values of a quiescent single BH with mass
$M_{\rm fin}=0.965M$ and dimensionless spin
$j_{\rm fin}=0.688$. We include in the upper panel
of Fig.~\ref{fig:max_k1k2} a single spike where $K_1\approx
150~K_{\rm S}$. This spike, however, consists of a single
data point, one time step before the first common horizon
is found; we regard it as likely that this spike is
spurious and may already be encompassed inside a common
horizon which the code simply failed to compute. We have decided
to still include this spike as a highly conservative upper
limit for the maximal $K_1$ realized in our BH simulations.

\noindent {\bf \em Conclusions.}
Our results are very clear. They show that in four-dimensional spacetimes describing realistic collapse configurations or dynamical BH spacetimes, curvature never grows too large during the dynamics. In other words, without carefully tuning the initial data, it seems very difficult to dynamically enter in a regime that would not be described by the classical equations of motion. Classical remains classical.

\noindent
{\bf \em Acknowledgments.} 
V.C.\ is a Villum Investigator and a DNRF Chair, supported by VILLUM FONDEN (grant no.~37766) and by the Danish Research Foundation. V.C.\ acknowledges financial support provided under the European
Union's H2020 ERC Advanced Grant ``Black holes: gravitational engines of discovery'' grant agreement
no.\ Gravitas–101052587.
This project has received funding from the European Union's Horizon 2020 research and innovation programme under the Marie Sklodowska-Curie grant agreement No 101007855.
We acknowledge financial support provided by FCT/Portugal through grants 
2022.01324.PTDC, PTDC/FIS-AST/7002/2020, UIDB/00099/2020 and UIDB/04459/2020.
This work has been supported by
STFC Research Grant No. ST/V005669/1
``Probing Fundamental Physics with Gravitational-Wave Observations''.
This research project was conducted using computational resources at the Maryland Advanced Research Computing Center (MARCC).
We acknowledge support by the DiRAC project
ACTP284 from the Cambridge Service for Data Driven Discovery (CSD3)
system at the University of Cambridge
and Cosma7 and 8 of Durham University through STFC capital Grants
No.~ST/P002307/1 and No.~ST/R002452/1, and STFC operations Grant
No.~ST/R00689X/1.
The authors acknowledge the Texas Advanced Computing Center (TACC) at The
University of Texas at Austin
and the San Diego Supercomputer Center for providing HPC resources that have contributed
to the research results reported within this paper through
NSF grant No.~PHY-090003. URLs: \url{http://www.tacc.utexas.edu}, \url{https://www.sdsc.edu/}.

\bibliography{main}

\end{document}